\documentclass{JINST}

\title{Druid, event display for the linear collider}

\author{Manqi RUAN$^a$\thanks{Corresponding author.}, Vincent BOUDRY$^a$, Gabriel MUSAT$^a$, Daniel JEANS$^a$, Jayant PANDE$^b$\\
\llap{$^a$} Laboratoire Leprince-Ringuet, \\
	\'Ecole polytechnique, CNRS/IN2P3, Palaiseau, 91128, France\\
\llap{$^b$} Department of Physics, Indian Institute of Technology, 
	\\208016 Kanpur, India\\

  E-mail: \email{Manqi.RUAN@llr.in2p3.fr}}

  \abstract{ 

	Druid is a dedicated event display designed for the future $e^{+}e^{-}$ linear colliders.
	Druid takes standard linear collider data files and detector geometry description files as input, it can visualize both physics event and detector geometry.
	Many displaying options are provided by Druid, giving easy access to different information. 
	As a versatile event display, Druid supports all the latest linear collider detector models, Silicon Detector and International Large Detector, as well as the calorimeter prototypes operated in the CALICE test beam experiments. 
	It has been utilized in many studies such as the verification of detector geometry, analysis of the simulated full events and test beam data as well as reconstruction algorithm development and code debugging. 

  }


\begin{document}

\section{Introduction }

A future $e^{+}e^{-}$ colliders with center-of-mass energy at the TeV level will play a key role in understanding electroweak symmetry breaking and physics beyond the standard model~\cite{ILC, CLIC}. 
To construct a detector that fulfils the physics requirements at a linear collider, pioneering prototypes equipped with new technologies have been constructed and large data sets have been collected in cosmic ray and test beam experiments~\cite{CALICEetal}. 
Full detector simulation~\cite{Mokka, SIDSimu}, reconstruction algorithms~\cite{JCPaper, PFA} are being developed and bench mark physics channels~\cite{LoI, SiDLoI} are being analyzed.
Druid (Display Root module Used for lInear collider Detector) was developed to support these studies. 

Following the idea of Particle Flow Algorithm~\cite{JCPaper, PFA}, ultra-high granularity calorimeters were employed in linear collider detector design, with which, by separating and measuring each jet particle in the most suited sub-detector, a very good jet energy resolution can be achieved.
Nowadays, with the development of micro-electronics, the granularity of the designed calorimeters for the linear collider are revolutionary increased comparing to previous experiments. 
For example, in the constructed physics prototypes of CALICE collaboration, the Silicon-Tungsten electromagnetic calorimeter holds 10k channels in a cube with side length of 20~cm~\cite{CALICEEcal}, the number of channels is one eighty of the CMS electromagnetic calorimeter~\cite{CMSEcal}. 
Both prototypes of digital and semi-digital hadron calorimeter have over half a million channels, leading to a world record of number of calorimeter channels in experimental physics~\cite{CALICEDHCAL, CALICESDHCAL}. 
The ultra-high granularity calorimeter provides an unprecedented level of details for the recorded shower, enables new approaches for shower analysis and jet reconstruction algorithm development. 

Besides the global event topology and detector geometry, Druid emphasizes at the high precision and flexible display.
The event/shower detail can be tagged through the zoom and rotation option.
Many display options are defined in Druid Graphic User Interface, providing easy access to different event information.
Seveal examples can be found in later sections. 
In this paper we present the performance, the dependencies, the display objects and options of Druid. 
At the end of this paper, we demonstrate two examples on using Druid to debug reconstruction algorithms. 

\section{Druid dependencies: LCIO, GDML and TEve}

Druid serves as a bridge between the displayed TEve~\cite{MatevPaper} objects and the information stored in LCIO data files and GDML geometry description files.
TEve is a framework for object management, providing hierarchical data organization, object interaction, and visualization through a ROOT~\cite{root} Graphical User Interface.
It is intensively used in LHC event displays.
LCIO~\cite{lciohome} is the standard linear collider data format while the GDML~\cite{gdmlpage} is an XML based geometry description format used to exchange geometry data between applications.
Druid has been optimized to have the minimal dependencies: only ROOT (version 5.28.00 or higher) and LCIO are requested.
Druid has been integrated into the ilcsoft~\cite{ilcsoft}.

\begin{figure}
\centering
\includegraphics[width=0.95\columnwidth]{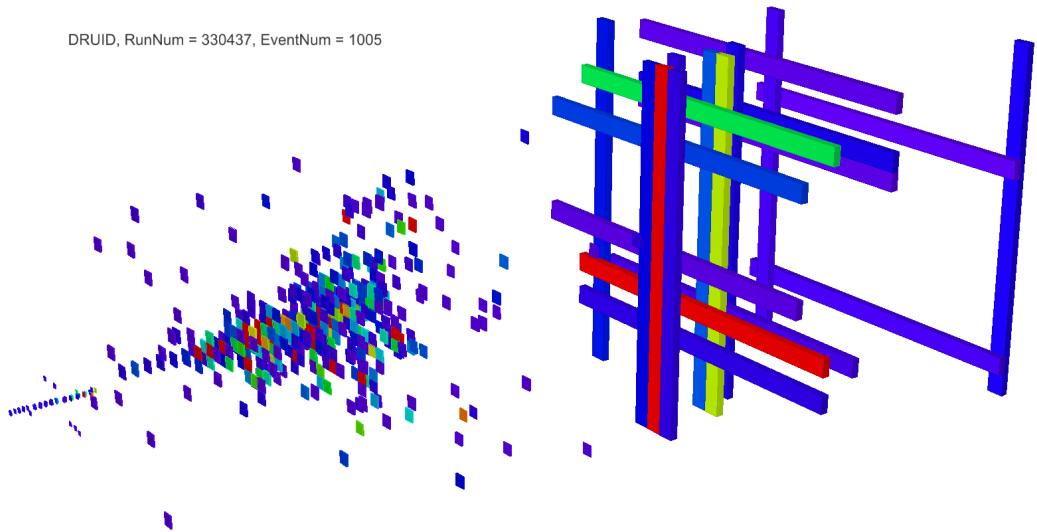}
\caption{40 GeV $\pi^{+}$ shower recorded at CALICE test beam experiment~\cite{CALICEetal}. The experimental setting up included the prototypes of an electromagnetic calorimeter (ECAL, with $1\times{1} cm^{2}$ cells), a hadronic calorimeter (HCAL with $3\times{3} cm^{2}$ cells) and a tail catcher (TCMT, $5\times{100} cm^{2}$ long strips). Their hits are displayed with accordingly sizes and colored to the hit energy. This image shows also the misalignment between ECAL and HCAL prototypes}
\label{CALICETBevt}
\end{figure}

\begin{figure}
\centerline{\includegraphics[width=0.95\columnwidth]{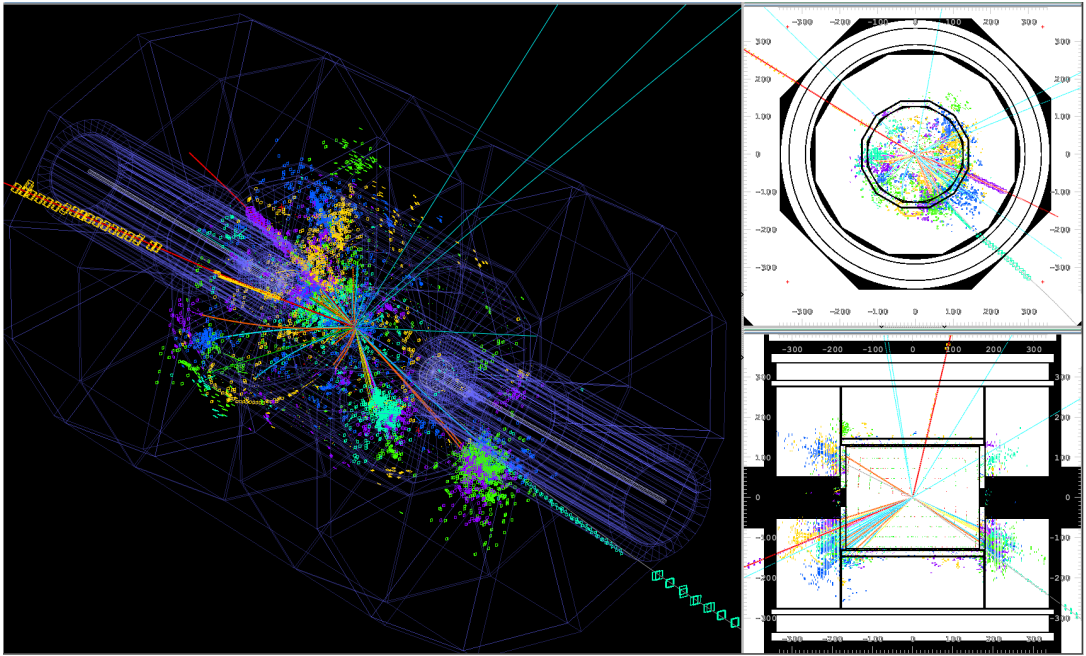}}
\caption{One TeV ttH event at the Compact Linear Collider~\cite{CLIC} Silicon Detector. One such event weight approximately 1~MB in the LCIO data format}
\label{TTH}
\end{figure}

Druid has been intensively tested on the fully simulated/reconstructed data at International Large Detector~\cite{LoI} (ILD) and Silicon Detector~\cite{SiDLoI} (SiD) and CALICE~\cite{CALICEetal} test beam data, see Fig.~\ref{CALICETBevt},~\ref{TTH}.
Testing on a 2.8 GHz laptop, it takes about 5 second to launch Druid at any data file.
The time needed to display an event is dependent on its data size, for example, about 3 second is required to display a One TeV ttH event at Compact Linear Collider such as the one on Fig.~\ref{TTH}.




\section{Event objects}

Event information is stored in different collections in a LCIO file. 
These collections can be classified into MCTruth level, Digitization level and the reconstruction level.
Accordingly, Druid defines the TEve objects and groups them following the same hierarchy. 
The correspondence is summarized in Table~\ref{tab1}.
\begin{table}
\caption{LCIO Collections versus corresponding TEve object}
\begin{center}
\begin{tabular}{ c | c | c }
\hline
Level & LCIO Collection & TEve Object	\\
\hline
		& MCParticle          &	TEveTrack			\\
MCTruth	& SimuCalorimeterHit  &	TEveBox				\\
		& SimTrackerHit		  &	TEvePointSet		\\
\hline
Digitization	& CalorimeterHit	&	TEveBox				\\
				& TrackHit			&	TEvePointSet		\\
\hline		
				& TrackAssignedHit		&	TEvePointSet 	\\		
Reconstruction	& Vertex				&	TEvePointSet	\\		
				& ClusterHit			&	TEveBox			\\		
				& ReconstructedParticle &	TEveTrack		\\		
\hline
\end{tabular}
\end{center}
\label{tab1}
\end{table}
The TEveTracks corresponding to MCParticle and ReconstructedParticle collections are organized into groups according to their particle type, while low energy objects are grouped for easy masking.  
For the detector hits collections, the TEveBoxs and TEvePointSets are divided into groups according to the subdetectors.
The size, color and style of the TEve objects can be defined on various information. 
For example, TEveTracks can be colored by particle type, and TEveBoxs corresponding to CalorimeterHit can be colored to hit energy, time, or the type of particle that induces this hit, see Fig~\ref{coloroption}.

For the detector geometry, the GDML file can be written by the simulation software~\cite{Mokka, SIDSimu}.
It records all the geometry information of the simulation: the size, the material, the orientation and the shape of every volume.
Druid displays each volume as a polyhedron whose color and transparency are determined by its material, allows a detailed verification on the detector geometry. 
The later release of Druid includes the GDML files for the five most recent full detector concepts as well as the CALICE test beam prototypes.

\begin{figure}
	\centering
	\includegraphics[width=0.55\columnwidth]{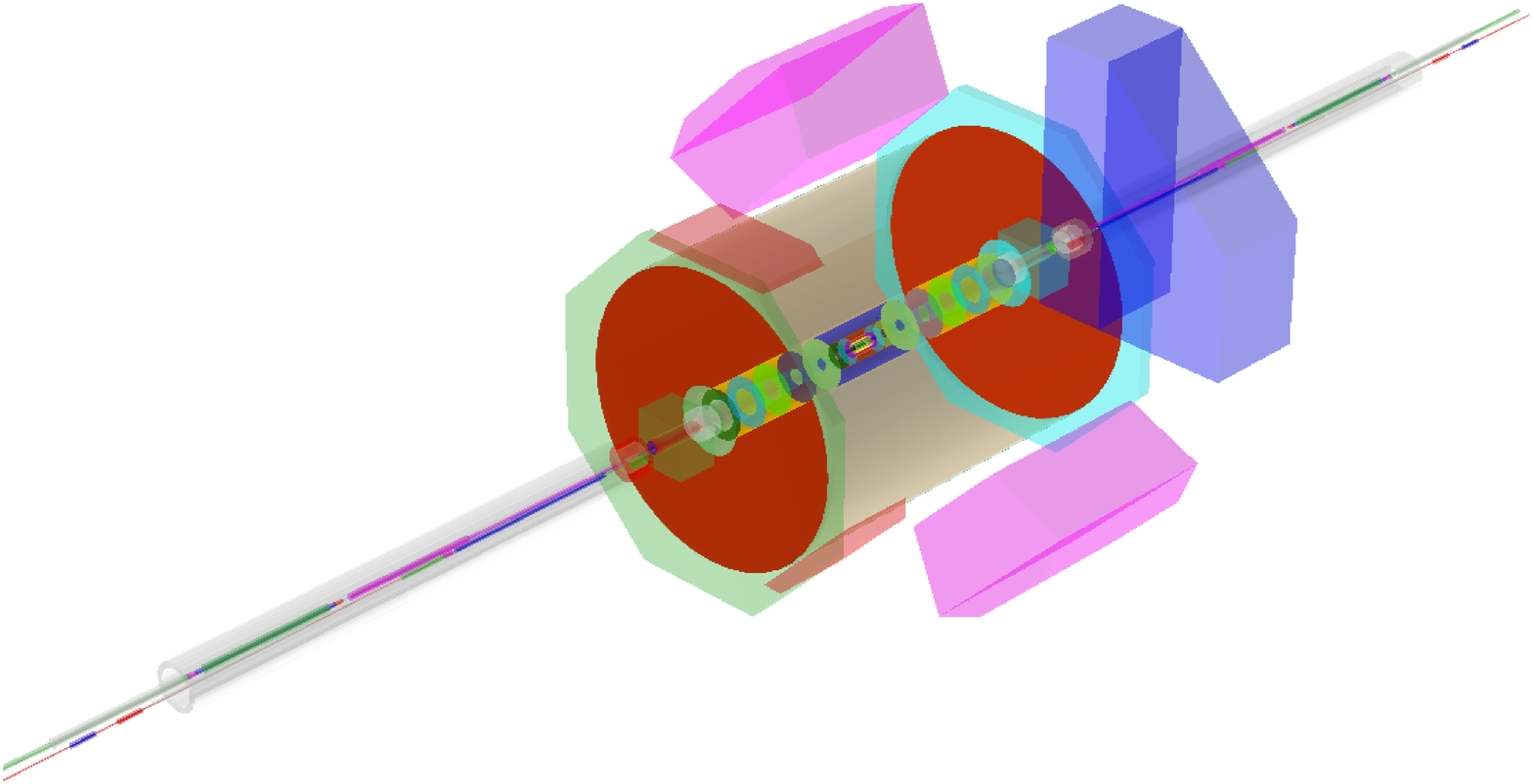}
	\hspace{0.1in}
	\includegraphics[width=0.40\columnwidth]{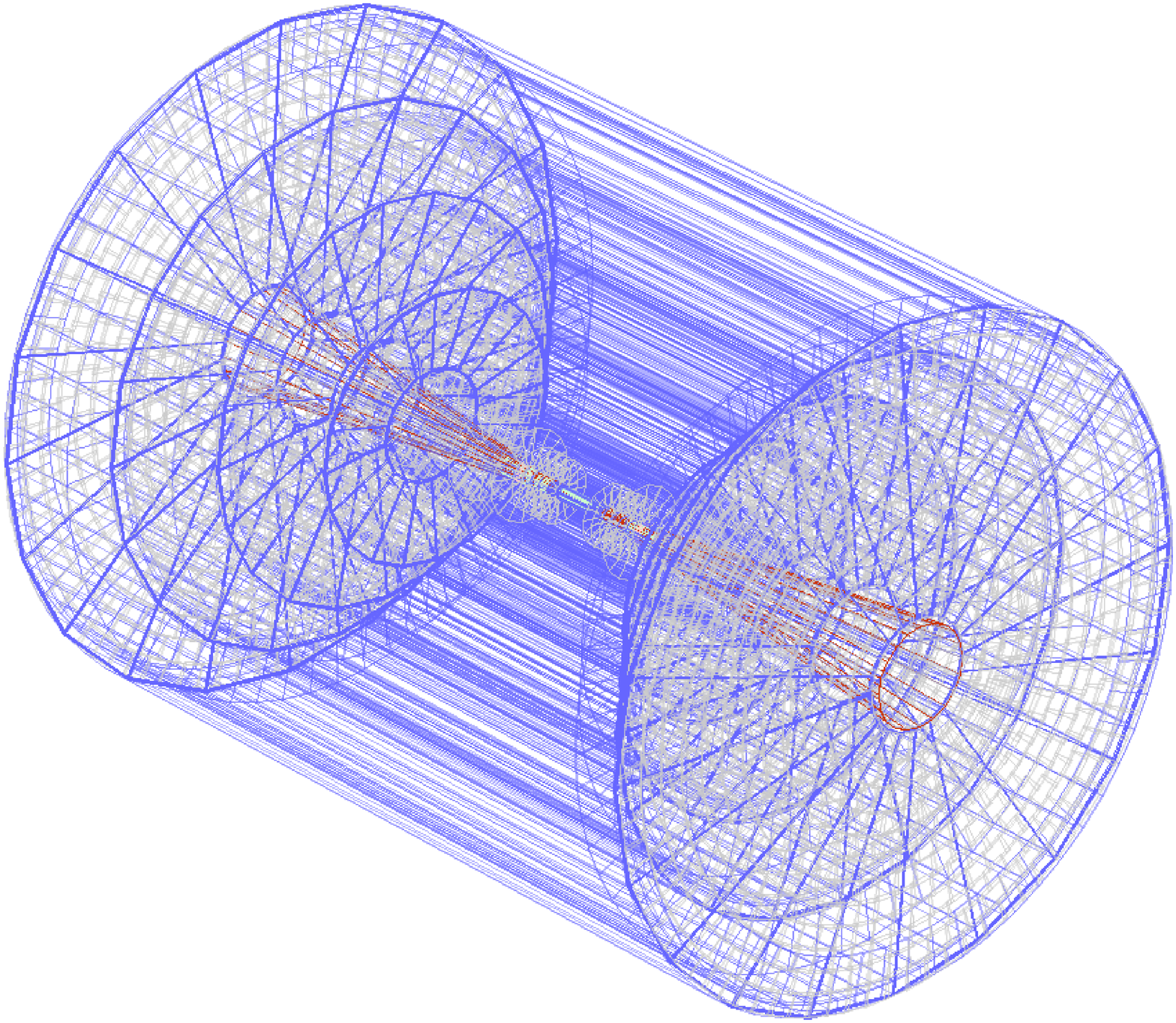}
	\caption{Control the level of details of geometry display with hide/dismount option and display depth: More details are displayed with higher display depth. \textbf{Left:} International Large Detector with the yoke, the coil and part of the calorimeter dismounted, lowest display depth \textbf{Right:} Tracking System of Silicon Detector at the second lowest display depth}
	\label{Dismount}
	\end{figure}

The full detector geometry is usually too detailed for a display focus on the event information.
Two options have been employed in Druid to reduce the workload of geometry display.
First, any ``unwanted" volume can be hidden, see Fig.~\ref{Dismount}.
Secondly, Druid employs the ``display depth", an global parameter referring to the hierarchy of geometry description in the GDML file, to interactively mask the geometry details. 
For the users focusing on the event information, the lowest display depth (the default), at which the contour of sub detectors are displayed, is usually sufficient.

	\section{Display options}

	\subsection{Options inherited from TEve}

\begin{figure}
	\centerline{\includegraphics[width=0.95\columnwidth]{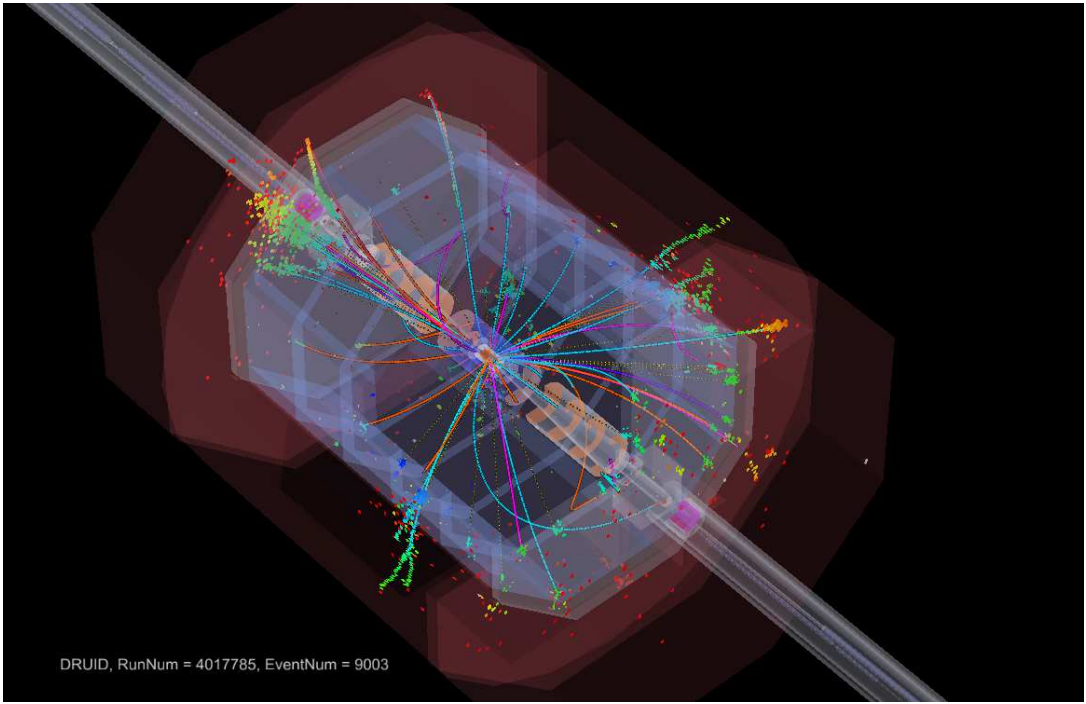}}
	\caption{500GeV ttbar event at ILD: calorimeter hits are colored according to the time of deposit}
	\label{ttb_cutview}
	\end{figure}

\begin{figure}
	\centering
	\includegraphics[width=0.95\columnwidth]{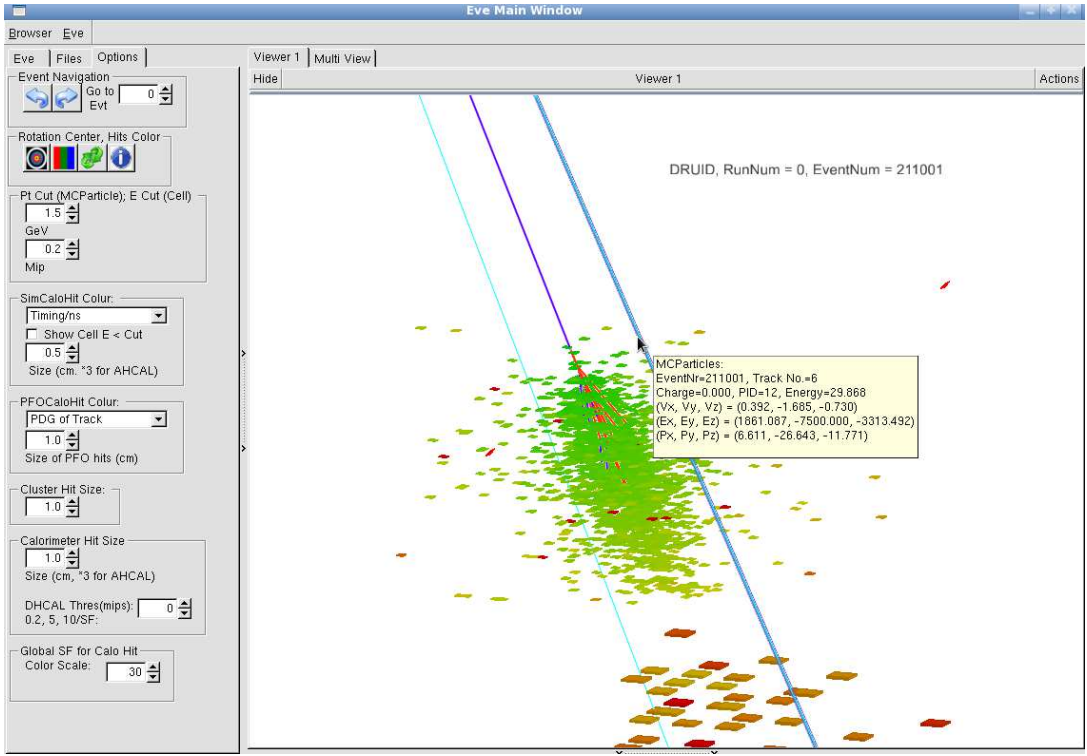}
	\caption{A simulated $\tau$ jet ($\tau^{-} \rightarrow e^{-} + \bar{\nu_{e}} + \nu_{\tau} $) of an 500~GeV $\tau\tau$ event at ILD. The Calorimeter Hits are colored according to their deposit time. The text attached to the $\bar{\nu_{e}}$ has been printed on the display, showing the basic information such as its particle type and 4-momentum. The option panel is shown in the left part of this image.}
	\label{siduds}
	\end{figure}

	Druid inherits many display options from TEve with the hot key access, such as the zoom, the rotation, return to the original orientation and scale as well as the black \& white background color switch.
	To focus on the inner part of the display, a cut away view can be used to removes part of the display. 
	One example is given in Fig.~\ref{ttb_cutview}, where one eighth of the detector is removed.
	TEve allows to attach text information on each displayed object.
	Picking an object with the censor, the attached text is printed in the display, see Fig.~\ref{siduds}.

	Besides these options, many interactive actions can be accessed at the ROOT GUI interface.
	The interface is divided into three pages: the file page, the eve page and the Druid option panel.
	The file page browses the file directory.
	The second page browsers all the generated TEve objects: displayed or hidden.
	For any TEve object, its display/hide status can be switched individually or by groups.
	Druid remembers the status of display/hide by the name of the collections when navigating to a new event. 
	Options as changing illumination setting, tuning geometry display depth, setting reference point/frame are also available in the second page.


	\subsection{Options defined in Druid}

	The third page includes the options defined by Druid, for example the event navigation, the object color/size setting and the cut parameters adjusting.
	There are also several options using buttons:

	(1). Select rotation center.

	(2). Regenerate the color: generate another color according to the object index.
	Here the index means the order of a given object in its collection, for example the clusters.

	(3). Switch between scenarios: the minimal scenario which ignores all the intermediate reconstructed objects such as digitized hits, tracks and clusters and the maximal scenario that displays every possible collection. 
	To give fastest performance, the minimal scenario is set as the default.

	To accelerate the speed of event display as well as to focus on interesting physics information, Druid defines several cuts with interactively adjustable parameters, for example the cut on the minimal transverse momentum on the MCParticle list and the cut on the energy of calorimeter hits.

\begin{figure}
	\centerline{\includegraphics[width=1.0\columnwidth]{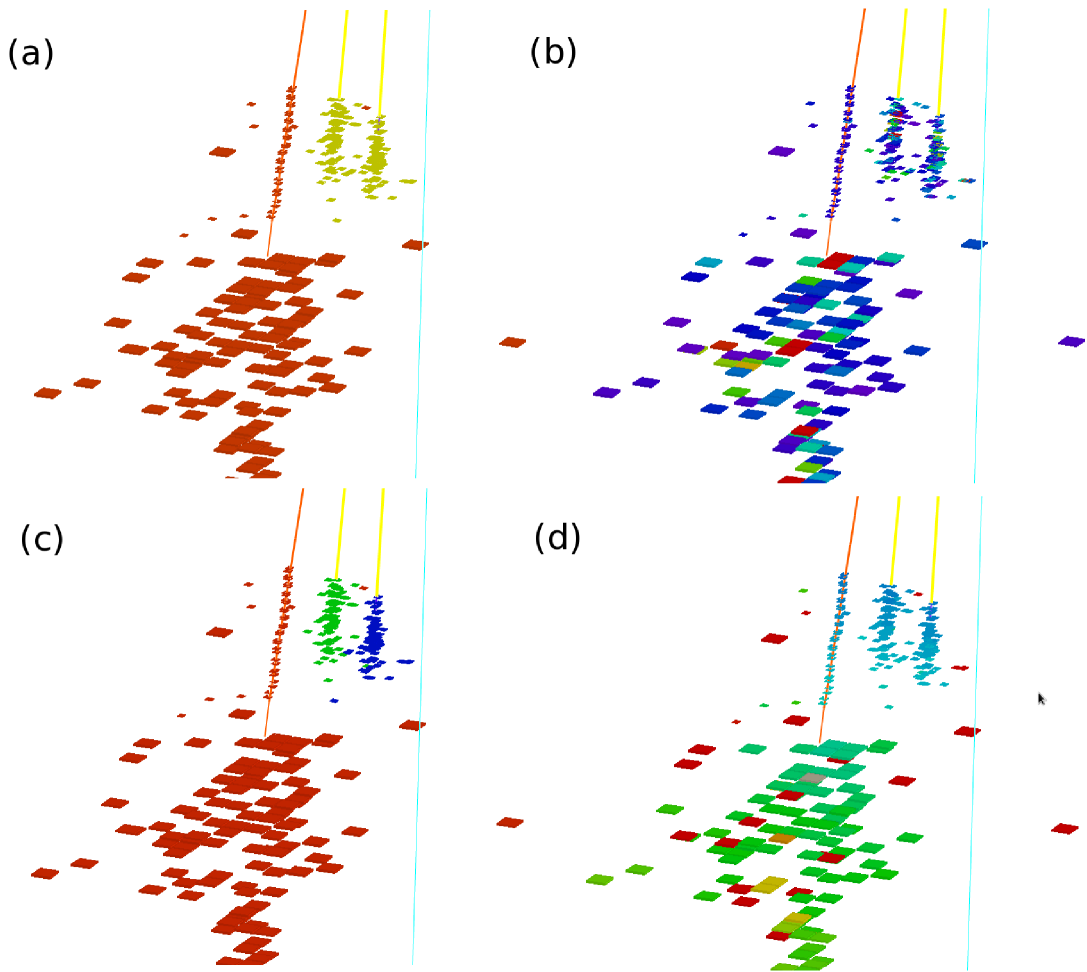}}
	\caption{Simulated $\tau$ jet ($\tau \rightarrow \nu + \pi^{0} + \pi^{+}$) at the calorimeter. Hit color are defined according to: 
		(a), Type of particle that induces the hit; (b), Energy of the hit; (c), Index of particle; (d), Deposit time of the hit.
		}
	\label{coloroption}
	\end{figure}

	As discussed in the introduction, the display of calorimeter hits is of special importance for the event display of linear collider. 
	Three different types of calorimeter hits are defined in LCIO: the simulated, digitized/recorded and clustered calorimeter hits. 
	The default size and orientation of calorimeter hits are set corresponding to different detector geometry concepts.
	The hit size can be changed independently for each type and the hit color can be specified according to different information. 
	The simulated hits can be colored to the energy, the type or index of the particle induces this hit, the time, or a uniform color.
	Fig.~\ref{coloroption} shows a simulated $\tau$ jet displayed with different color option.
	The digitized calorimeter hits are colored to the energy, and the clustered hits are colored according to the index of the cluster.
	A global factor can be adjusted to scale the calorimeter hit color when it is colored to the hit energy or the deposit time;
	Once colored to the index, the color can be regenerated to give a better separation to nearby hits induced by the same kinds of particle, see Fig.~\ref{coloroption}(c).
	Once a cut or a hit size/color configuration has been adjusted, the name and statistics of the affected collections are printed on the terminal.

	\section{Example application: debugging reconstruction code}

	One of the most important applications for Druid is the debugging of reconstruction code.
	Here we demonstrate two examples with PandoraPFA, the most successful reconstruction Particle Flow Algorithm developed for linear collider.

\begin{figure}
	\centerline{\includegraphics[width=1.0\columnwidth]{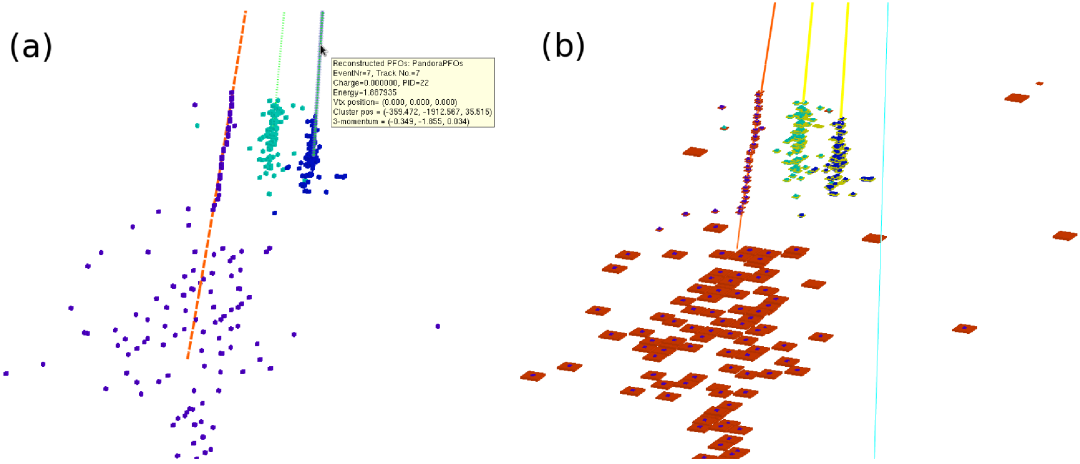}}
	\caption{A $\tau$ jet ($\tau \rightarrow \nu + \pi^{0} + \pi^{+}$) reconstructed with PandoraPFA.
		(a) Reconstructed objects: ReconstructedParticle and corresponding cluster.
		(b) MCTruth and reconstructed objects: SimCalorimeterHit, MCParticle, ReconstructedParticle and its cluster.  
		}
	\label{Reccompare}
	\end{figure}

	The first example is the reconstruction of a $\tau$ jet: we recall the same $\tau$ jet as in Fig.~\ref{coloroption}.
	Fig.~\ref{Reccompare}(a) shows the reconstructed particle and their corresponding cluster, where two photons and one pion has been reconstructed.  
	The cluster hits are displayed as cubes with 5~mm side length.  
	In Fig.~\ref{Reccompare}(b), Druid overlays the reconstructed objects with MCTruth objects: SimCalorimeterHits and MCParticle.
	The blue straight line indicates a neutrino generated from $\tau$ decay. 
	The SimCalorimeterHits are colored according to the particle type, red for pions and yellow for photons.
	Most of the SimCalorimeterHits has been attached to the reconstructed particle. 
	Therefore, for this $\tau$ jet, the output of the reconstruction algorithm agrees with the MCTruth information. 
	Reading the attached text information, further comparison on the reconstructed energy and MCTruth energy for each particle is accessible. 
	
	The second example is a 100~GeV $\pi$ shower. 
	Fig.~\ref{Reccompare}(a) shows the simulated detector hits. 
	The $\pi$ hits into the calorimeter, create a hadronic cluster composed of two electromagnetic sub clusters and several sailing through tracks as well as a separated small cluster deposited by a backscattering charged particle. 
	The reconstructed objects is shown in Fig.~\ref{Reccompare}(b), where PandoraPFA divided these hits into four clusters: 
	The leading cluster is associated with the track, reconstructed as a charged particle with the energy equal to track momentum (the total energy). 
	These remaining three clusters are reconstructed as nearby neutral particles, create a significant amount of double counted energy. 
	Actually, Fig.~\ref{Reccompare}(b) shows a typical Particle Flow Algorithm double counting. 
	To reduce such kind of confusions is the main challenge for Particle Flow Algorithm development. 

	\begin{figure}
	\centerline{\includegraphics[width=1.0\columnwidth]{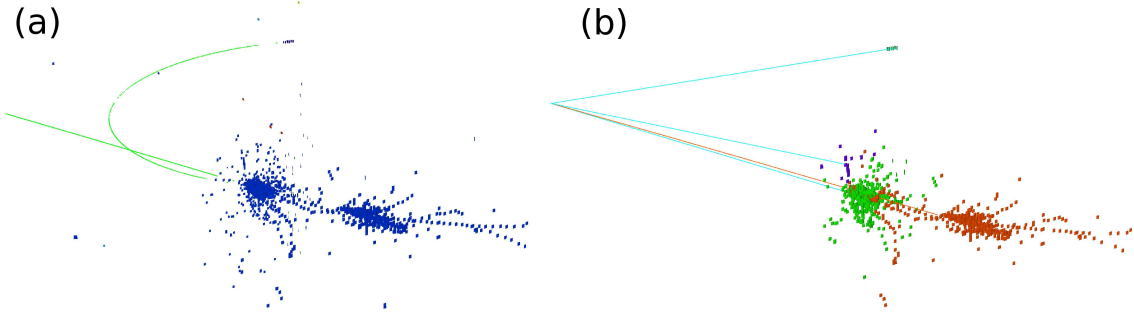}}
	\caption{A $\pi$ shower reconstructed with PandoraPFA.
		(a) MCTruth objects: SimCalorimeterHits and SimTrackerHits.
		(b) Reconstructed object: ReconstructedParticles and corresponding clusters.
	}
	\label{PionReccompare}
	\end{figure}

\section{Summary}

Druid, a dedicated event display for linear collider has been developed.
For the event data, Druid not only displays the global event topology but also provides close view to the event/shower details, with the options to emphasize on different information. 
Reading GDML file written by simulation software, Druid can display all detailed simulated detector geometry with practical options to control the level of details and to browser the geometry. 
It has been heavily used in geometry verification, data analysis and reconstruction algorithm development.

\acknowledgments

We are grateful to Norman Graf, for the suggestion of using GDML file; to Henri Videau, for his suggestions on the display style setting, to Jean-Claude Brient, for his continuous support in this project.
Special Thanks goes to Matevz Tadel and Alja Tadel, for all their support and discussions.
The research leading to these results has received funding from the European Commission under the FP7 Research Infrastructures project AIDA, grant agreement no. 262025.



\begin{thebibliography}{9}

\bibitem{ILC} J. Brau, \emph{et al.} \emph{International linear collider reference design report} \emph{CERN Report} (2007) CERN-2007-006 ILC-REPORT-2007-001
\bibitem{CLIC} M. Battaglia, \emph{et al.} \emph{Physics at the CLIC Multi-TeV Linear Collider: report of the CLIC Physics Working Group} \emph{CERN Report} (2004) CERN-2004-005 (\emph{Preprint} hep-ph/0412251)
	\bibitem{CALICEetal} I. Laktineh \emph{CALICE results and future plans} PoS(ICHEP 2010) 493, see also https://twiki.cern.ch/twiki/bin/view/CALICE/WebHome
	\bibitem{Mokka}	http://polzope.in2p3.fr:8081/MOKKA/
	\bibitem{SIDSimu} http://lcsim.org
	\bibitem{JCPaper} J-C. Brient, \emph{Particle Flow Algorithm and calorimeter design J. Phys.: Conf. Series} \textbf{160} (2009) 012025
	\bibitem{PFA} M.A. Thomson, \emph{Particle Flow Calorimetry and the PandoraPFA Algorithm Nucl. Instrum. Meth. A} \textbf{610} (2009) 25-40
	\bibitem{LoI} T. Abe, \emph{et al.} \emph{The International Large Detector: Letter of Intent} \emph{DESY Report} (2010) DESY-2009-87
	\bibitem{SiDLoI} H.~Aihara {\it et al.}  [SiD Collaboration], \emph{SiD Letter of Intent}, SLAC-R-944
	\bibitem{CALICEEcal} J.Repond, \emph{et al.} 2008 JINST 3 P08001	
	\bibitem{CMSEcal} The CMS Collaboration, 2006, CMS-TDR-008-1
	\bibitem{CALICEDHCAL} B.Bilki, \emph{et al.} 2009 JINST 4 P10008
	\bibitem{CALICESDHCAL} M. Bedjidian, \emph{et al.} 2011 JINST 6 P02001
	\bibitem{MatevPaper}
        M. Tadel, \emph{Overview of Eve - the Event Visualization Environment of the ROOT} \emph{J. Phys.: Conf. Series.} \textbf{219} (2010) 042055
	\bibitem{lciohome} F. Gaede, \emph{et al.} \emph{LCIO-A persistency framework for linear collider simulation studies} (2003) \emph{LC-TOOL-2003-053} 
	\bibitem{gdmlpage} R. Chytracek, J. McCormick, W. Pokorski, G. Santin, \emph{Geometry Description Markup Language for Physics Simulation and Analysis Applications} \emph{IEEE Trans. Nucl. Sci.} Vol.{\bf 53} (2006) Issue: 5, Part 2, 2892-2896
	\bibitem{root} R. Brun~R and F. Rademakers, \emph{ROOT - An Object Oriented Data Analysis Framework, Proceedings AIHENP'96 Workshop, Lausanne, Sep. 1996} \emph{Nucl. Inst. Meth. in Phys. Res. A} {\bf 389} (1997) 81-86
	\bibitem{OpenGL} http://www.opengl.org/
	\bibitem{G4paper}  S. Agostinelli, \emph{et al.} \emph{Geant4 a simulation toolkit} Nucl. Instrum. Meth. A {\bf 506} (2003) 250-303
	\bibitem{ilcsoft} http://ilcsoft.desy.de
	\bibitem{ilcinstallpage} http://ilcsoft.desy.de/portal/software\_packages/ilcinstall/


	\end{thebibliography}
	\end{document}